\documentclass[twocolumn,trackchanges]{aastex7}

\usepackage{amsmath}
\usepackage{hyperref}
\usepackage{float}

\begin{document}

\title{Comparing Results from Two Uniform Phase Curve Surveys}

\author[orcid=0009-0008-3326-9715,sname='Emeline Decocq']{Emeline Decocq}
\affiliation{Kapteyn Astronomical Institute, University of Groningen, Groningen, the Netherlands}
\email{e.m.m.decocq@rug.nl}  

\author[orcid=0009-0001-4487-7299, sname='Mark Swain']{Mark Swain} 
\affil{Jet Propulsion Laboratory, California Institute of Technology, Pasadena, USA}
\email{mark.r.swain@jpl.nasa.gov}

\author[orcid=0000-0003-4987-6591, sname='Lisa Dang']{Lisa Dang}
\affiliation{Department of Physics and Astronomy and the Waterloo Centre for Astrophysics, University of Waterloo, Waterloo, Canada}
\email{lisa.dang@uwaterloo.ca}

\author[orcid=0000-0002-5741-3047,sname='David R. Ciardi']{David R. Ciardi}
\affil{NASA Exoplanet Science Institute, MS 100-22, California Institute of Technology, Pasadena, USA}
\email{ciardi@ipac.caltech.edu}

\author[orcid=0000-0001-5966-837X,sname='Geoffrey Bryden']{Geoffrey Bryden}
\affil{Jet Propulsion Laboratory, California Institute of Technology, Pasadena, USA}
\email{Geoffrey.Bryden@jpl.nasa.gov}

\begin{abstract}

We present a comparison of the two most recent and comprehensive Spitzer phase curve studies -- \cite{dang2025} and \cite{swain2025} -- which report analyses of the Spitzer 4.5 $\mu{m}$ phase curves.  The studies employ different approaches for correcting instrument systematics and they also use different approaches for selecting the optimal exoplanet system parameters. To evaluate the level of consistency between the two studies, we compared the constraints on the ratio of planet-to-star radii ($\frac{R_{P}}{R_{\star}}$), eclipse depth ($\frac{F_{P}}{F_{\star}}$), phase curve amplitude (A), and phase curve offset ($\phi$). We find that the two studies produce similar results at the population level although results for individual planets can vary, especially for phase curve offset values. We examined the difference of planet system parameters to see if inconsistencies in individual planet results were due to data reduction methods or system parameter choices. We also examined whether the system parameters used by both studies were consistent with Kepler’s third law. During this comparison, we identified one case where stellar mass, planet semi-major axis, and orbital period did not follow Kepler's law even though the values were all compiled from the same publication. To assess whether this kind of discrepancy was recurrent, we recalculated the orbital periods using Kepler’s third law and compared them with the values listed in the NASA Exoplanet Archive (NEA). Our detailed analysis of archival system parameters strongly suggests that testing reported/selected parameters for consistency with Kepler's third law is worthwhile.


\end{abstract}



\section{Introduction}\label{sec:Introduction} 

During the extended mission phase, the Spitzer Observatory \citep{Werner2004} observed the thermal phase curves of more than 30 exoplanets with the IRAC instrument \citep{Deming:2020aa,dang2025}. These legacy Spitzer observations provide a unique resource for the statistical study of day-to-night heat transport in the atmospheres of gas giant planets undergoing powerful external radiation forcing.

Recognizing the wealth of multi-dimensional insights embedded in thermal phase curve, a number of studies have searched for trends in thermal phase curves properties \citep{schwartz2015,schwartz2017,zhang2018,bell2021,may2022,dang2025,swain2025} with the most recent studies \citep{bell2021,dang2025,swain2025} implementing uniform analysis methods to facilitate comparative planetology studies.

Spitzer IRAC observations are well known to suffer from substantial systematic errors, for which a number of methods correction methods have been developed (see discussion in \cite{dang2025}). The question of how inferred science parameters for individual dataset are impacted by a choice of selection methods has been examined for individual sources, most notable WASP-43b  \citep{stevenson2017,mendonca2018,morello2019,may2021,bell2021,murphy2023,dang2025,swain2025}.

In this study, we undertake the first catalog level comparison of results from the two most comprehensive phase curve studies \citep{dang2025,swain2025} hereafter refereed to as D25 and S25. While both the D25 and S25 studies are similar in that they undertake a uniform analysis of a substantial number of 4.5$\mu$m phase curves, the studies have three important differences. Firstly, D25 and S25 bring different interpretations to how the uniform analysis is defined. D25 uses a uniform criteria to select the optimal method for correcting instrument systematic (resulting in different targets in the sample being processed with different systematic correction methods) where as S25 uses single method for correcting instrument systematic that is applied uniformly to the sample. 

Secondly, D25 and S25 use a different approach for selecting the optimal exoplanet system parameters from among numerous measurements. In particular, they focus on constraining the phase curve, which requires a careful analysis of four key parameters; the planet-to-star radius ratio ($\frac{R_{P}}{R_{\star}}$), the secondary eclipse depth ($Depth_{eclipse}$), the phase curve amplitude (A), and the phase curve offset ($\phi$). Together, these parameters provide insights of the redistribution of heat across the planet’s atmosphere as well as the geometry of the system. The modeling of these parameters relies not only on the planetary radius but also on stellar properties.

Since a single phase curve cannot constrain the stellar properties and orbital ephemerides; D25 imposes a gaussian prior on these parameters by using an algorithm that selects system parameters from the NASA exoplanet archives (NEA)  (\cite{christiansen2025}) with the smallest uncertainty, regardless of whether the parameters are all from the same paper or from different papers which could result in mutually inconsistent system parameters. In contrast, S25 uses an algorithm which prioritizes identifying a set of self-consistent system parameters derived from a single paper. 

Thirdly, the D25 and S25 studies differ in scope. D25 is focused on trends for 4.5$\mu$m phase curves where as S25 is focused on a smaller subset of planets that have both 3.6$\mu$m and 4.5$\mu$m phase curves.

In what follows, we compare the D25 and S25 studies based on the mutual trends for the key astrophysical parameters reported by the studies. We also discern the possible origin for differences in results in the cases where the studies report findings that are in tension. And finally, we compare the orbital periods derived from our calculations with those reported in the Exoplanet Archive.


\section{Method} \label{sec:Method}

We compare the D25 and S25 results based on the estimates of four key physical exoplanet parameters; $R_{p}/R_{\star}$, where $R_{p}$ is the planet radius (in $R_{J}$) and $R_{\star}$ is the stellar radius (in $R_\sun$), the eclipse depth ($Depth_{eclipse}$) is the flux of the planet divided by the flux of the star (in ppm), $A$ is the phase curve amplitude (in ppm), and $\phi$ is the phase of the phase curve modulation relative to the midpoint of the exoplanet eclipse event (in degrees east of the substellar point). 

We will start by examining the radius and the eclipse depth. These two criteria are important to us because, Figure 3 in \cite{swain2025}, in the left-hand column and the third box up from the bottom, shows that the retrieval results are not correlated.  Therefore, we can focus on these two parameters. For $R_{p}/R_{\star}$, the D25 and S25 results can be directly compared. For the eclipse depth ($Depth_{eclipse}$) we transform the S25 representation into the D25 representation,

\begin{equation}
Depth_{eclipse} = \left(\frac{R_{p}}{R_{\star }}\right)^{2}\times \frac{F_{p}}{F_{\star }}
\label{eq:Fp_Fs}
\end{equation}

Similarly, the D25 phase curve amplitude, $A$, is calculated as half the distance between $F_{max}$ (the highest point) and $F_{min}$ (the lowest point). So we obtain the amplitude, using the following equation:

\begin{equation}
A = \frac{F_{max} - F_{min}}{2}
\label{eq:A}
\end{equation}
with $F_{\mathrm{max}}$ and $F_{\mathrm{min}}$ in ppm.
The phase curve phase, $\phi$, sign convention differed between D25 and S25; D25 define the phase curve offset as degrees east of the substellar point, whereas S25 defined it as the shift in orbital phase of the peak flux.
We multiply the S25 values by $-1$ so that the values can be directly compared as degrees east.

We also check the system parameter values used in D25 and S25 for self consistency by comparing the measured period $P$, to the derived period based upon Kepler's Law and the system parameters from the NEA, $a$, the stellar mass $M_{\star}$ and the planet mass $M_{p}$ using :

\begin{equation}
P_{Kepler} = \sqrt{ \frac{4 \pi^{2}  a^{3}}{G  (M_{\star} + M_{p})}}
\label{eq:P_Kepler}
\end{equation}
where $G$ is the gravitational constant, and since $M_{p} \ll M_{\star}$, $M_{p}$ can be neglected. We estimate the uncertainty in $P_{Kepler}$ with :

\begin{equation}
\sigma_{P_{Kepler}} = \sqrt{ \frac{9 \pi^{2} a} {G  M_{\star}} \sigma{a}^{2} + \frac{\pi^{2} a^{3}} {G  M_{\star}^{3}} \sigma{M}^{2}},
\label{eq:errorP1}
\end{equation}
or
\begin{equation}
\sigma_{P_{Kepler}} = \sqrt{ \left(\frac{1.5 \sigma{a}} {a }\right)^{2} + \left(\frac{0.5 \sigma{M_{\star}}}{M_{\star}}\right)^{2}}
\label{eq:errorP2}
\end{equation}

Given values for $P_{Kepler}$ and $\sigma_{P_{Kepler}}$, we construct a Keplerian Consistency Score (KCS) using,
\begin{equation}
\text{KCS} = \frac{\left| P_{\text{paper}} - P_{\text{Kepler}} \right|}{\sigma_{P_{\text{Kepler}}}}
\label{eq:KCS}
\end{equation}
This equation can be generalized as a Consistency Score (CS) to statistically compare the values of each parameter X between D25 and S25. This parameter is defined as the absolute difference between two measurements divided by the quadrature sum of their standard deviations, assuming the uncertainties are normally distributed and uncorrelated :

\begin{equation}
\text{CS} = \frac{\left| X_{D25} - X_{S25} \right|}{\sqrt{\sigma_{X_{D25}}^{2} + \sigma_{X_{S25}}^{2}}}
\label{eq:CS}
\end{equation}
This score quantifies the statistical consistency of the two measurements.

Additionally, we compared the planet equilibrium temperatures, $T_{eq}$ used by D25 and S25, for consistency with a planet equilibrium temperature based on system parameters, $T^{sys}_{eq}$, using, for S25, the following equation,

\begin{equation}
T^{sys}_{eq} = T_{\star} \sqrt{\frac{R_{\star}}{2 a}}  (1 - A_{B})^{\frac{1}{4}}
\label{eq:Teq1}
\end{equation}

For this equation, we assume that $A_{B}$ is equal to 0, as was done by both D25 and S25.

Values for the constants used in the equations are shown in Table~\ref{tab:constants} in the appendix. 
To assess if the derived parameters are consistent between the two papers (D25 and D25), we have plotted each parameter and compared them to unity (Figure ~\ref{fig:phys_parms}). In the next section, we discuss the results of the comparison between the two papers - including how the consistency of the parameters utilized in the calculations may affect the derived parameters

\section{Results} \label{sec:Results}

Here we summarize the comparison of the D25 and S25 studies in terms of physical parameters and the assumptions of the parameters used to describe the system.

\subsection{Phase Curve Parameters}\label{sec:Phase curve Parameters}
 
\begin{figure*}[t!]
\centering
\hspace{-0.25in} \\
\includegraphics[scale=1.1]{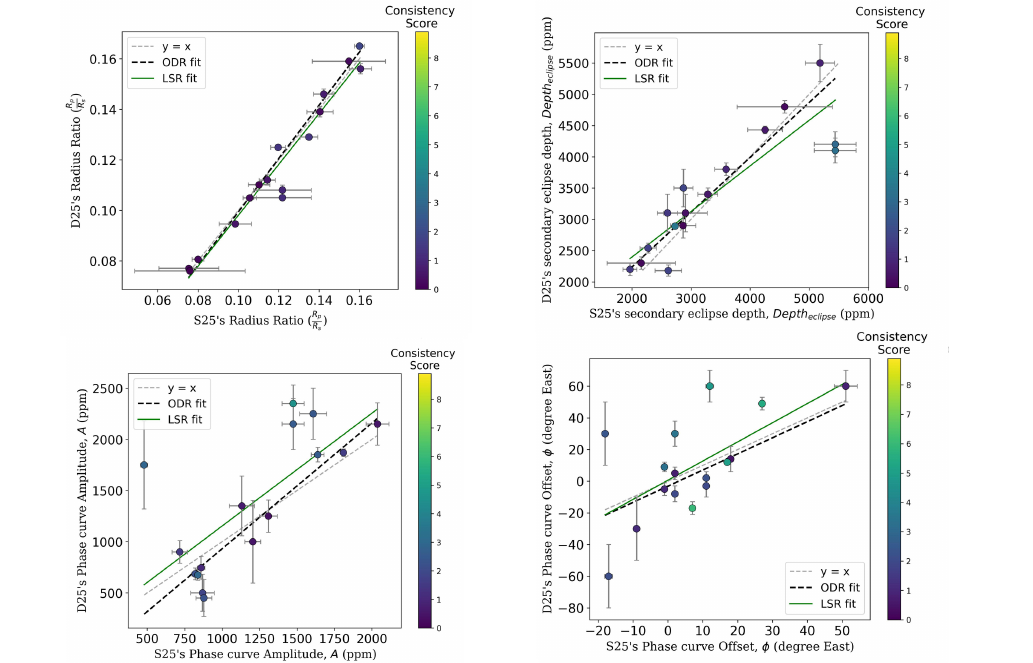}
\caption{A comparison of the most important phase curve parameters reported in the studies compared. Starting at the upper right and proceeding clockwise the panels compare the parameters: $R_{p}/R_{\star}$, $Depth_{eclipse}$ (in ppm), $A$ (in ppm), and $\phi$ (in degree East). We have overlaid the unity line along with, two linear fits, Orthogonal Distance Regression (ODR) and Least Squares regression (Ordinary Least Square).}
\label{fig:phys_parms}
\end{figure*}

In this section we first present the comparisons of the key exoplanet science parameters results reported in D25 and S25. The results are summarized in Table~\ref{tab:description} and Figure~\ref{fig:phys_parms} where points that are consistent with the unity slope line and not scattered indicate consistent results between the D25 and S25 studies.

To assess whether the data points are consistent with the unity slope line (black line), we applied two different fitting methods, the Orthogonal Distance Regression (ODR) and Ordinary Least Squares (OLS).
The ODR (black curve) method minimizes (taking into account measurement errors) orthogonal distances between data points and the fitted line. In order to have a better comparison, we also made an OLS adjustment (green curve), which minimizes the sum of squared differences between observed values and predicted values by the line. This method allows to obtain a unique regression equation that better represents the relationship between variables (but this time without taking into account the measurement error). One can therefore, thanks to these two lines, observe the influence of measurement uncertainties and thus in the case where the two lines are strongly separated, the uncertainties on the measurements have a noticeable effect on the regression.

The figure shows that, for $R_{p}/R_{\star}$ and the eclipse depth, the data are not scattered and both trend curves are close to the unit curve meaning that the results from both paper are globally consistent.

\begin{figure}[t!]
\centering
\includegraphics[scale=0.9]{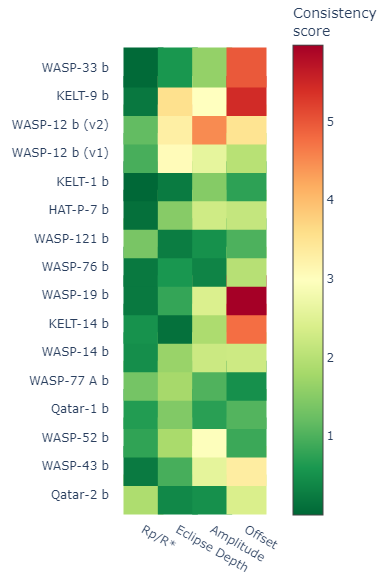}
\hspace{0.5in}
\caption{A comparison of the consistency score for $R_{p}/R_{\star}$, $Depth_{eclipse}$, $A$, and $\phi$ using the equation \ref{eq:CS} highlights that $R_{p}/R_{\star}$ and $Depth_{eclipse}$ are mostly consistent, whereas the phase curve offset shows the least consistency. The corresponding results are reported in Table \ref{tab:comp4}.} 
\label{fig:comparison_parameters}
\end{figure}

On closer inspection, however, we can see that some planets, particularly for parameters A and $\phi$, show data that deviate significantly from the unit slope line. In order to see these data more clearly, we can calculate the consistency score between these two papers. This will allow us to see, for each planet, the degree of difference between the D25 and S25 papers, to see if the data are indeed globally consistent with each other (Figure ~\ref{fig:comparison_parameters}). This matrix demonstrate the consistency score between the parameters from D25 and S25, as calculated using the equation \ref{eq:CS}. This score shows the distance between D25 and S25 data, a score lower than 1 $\sigma$ means that the data are consistent and the higher the score, the most probably inconsistent the data are.

Figure ~\ref{fig:comparison_parameters} provides a more detailed illustration of the disparities in values between the D25 and S25 data. It is evident that, for $R_{p}/R_{\star}$ all the data are consistent (with a score less than 2$\sigma$). And for the other parameters, most of the data are consistent with in total more than 70.83 \% with a consistency score less than 2.5$\sigma$.

However we can see that some planets have a high score indicating an inconsistency. For the eclipse depth we can see 2 planets with a consistency score higher than 3$\sigma$ : WASP-12 b and KELT-9 b (with respectively 3.3 $\sigma$ and 3.55$\sigma$). For this fit, WASP-12 b showed significant evidence for ellipsoidal variation so D25 has additionally included this effect in the phase curve model, however, in S25, no parameters for ellipsoidal variation were included, so this might explain the difference. For the amplitude, there are 5 planets with a consistency score higher than 2.5$\sigma$. Meanwhile for the phase curve offset, despite the fact that the trend curve is close to the unity slope line, 6 planets have a value superior than 2.5$\sigma$, ranging from 0.50 for WASP-77 AB to 5.97$\sigma$ for WASP-19 b. Thus, most of the planet parameter values are consistent but there are cases of inconsistencies in values between the two studies. These inconsistencies could be due to differences in data reduction or differences in system parameter selection.

\subsection{System Parameters}\label{sec:System Parameters}

\begin{figure}[ht]
\centering
\includegraphics[scale=0.9]{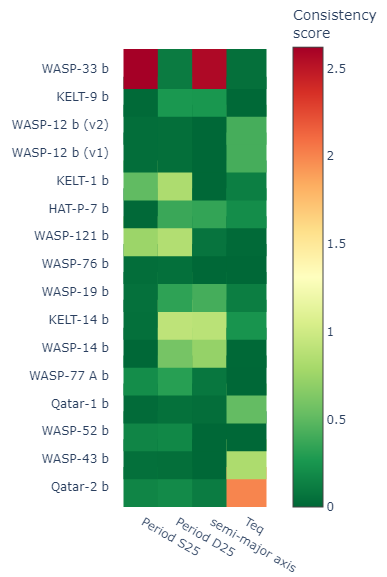}
\hspace{0.5in}
\caption{A comparison of the consistency scores for three system parameters is presented. Equation \ref{eq:KCS} was used to compare the periods from S25 with the periods calculated using the equation \ref{eq:P_Kepler} (Period S25, in days) and the periods from D25 with those calculated using the equation \ref{eq:P_Kepler} (Period D25, in days). Equation \ref{eq:CS} was used to compare the semi-major axes from D25 and S25 (semi-major axis, in AU) and the equilibrium temperatures from D25 and S25 (Teq, in K). The corresponding results are reported in Table \ref{tab:comp3}.} 
\label{fig:comparison_parameters2}
\end{figure}

As documented in the respective publications, D25 and S25 use different values for the system parameters that describe the host star, exoplanet, and planetary orbit. The system parameter values serve as input to the eclipse fitting and thus there is a possibility that differences in system parameter assumptions between D25 and S25 might cause differences in results. To explore this possibility, we made comparisons of selected system parameters used by D25 and S25 (Figure~\ref{fig:comparison_parameters2}). 
The measured orbital periods were compared to the orbital periods derived from the system parameters via Kepler's Law for each of the two papers D25 and S25. Subsequently, we used equation \ref{eq:CS} to assess the differences in the semi-major axis and the equilibrium temperature between D25 and S25.

Figure~\ref{fig:comparison_parameters2} shows in more detail the comparison of the selected system parameters of D25 compared to the ones from S25. We compared the results to assess whether the data used by D25 and S25 are consistent or differ, and, in the latter case, to investigate the reasons for these differences. 
From this figure, we can see that, for most cases, the results for the semi-major axis and the period fall below 1$\sigma$, indicating that they are most likely consistent. However, there is one planet, WASP-33 b, with a consistency score of 2.62$\sigma$ for the semi-major axis. This discrepancy requires further investigation, which will be discussed later. If we look, now, at the equilibrium temperature, we can see that all the planets have consistencies scores less than 2$\sigma$, showing no discrepancies between the studies. 

\section{Discussion}\label{sec:Discussion}

In this section the results of the study are discussed. The previous sections showed that for most planets, the retrieved parameters are broadly consistent between D25 and S25 (see Figure~\ref{fig:phys_parms}). However there are a few planets that have some discrepancies (up to 5.97$\sigma$) and we seek to determine the likely origin of differences in results. To explore the source of these discrepancies, we begin by analyzing the system parameters to assess whether their choice could account for the observed differences.

\begin{figure}[ht]
\includegraphics[scale=0.47]{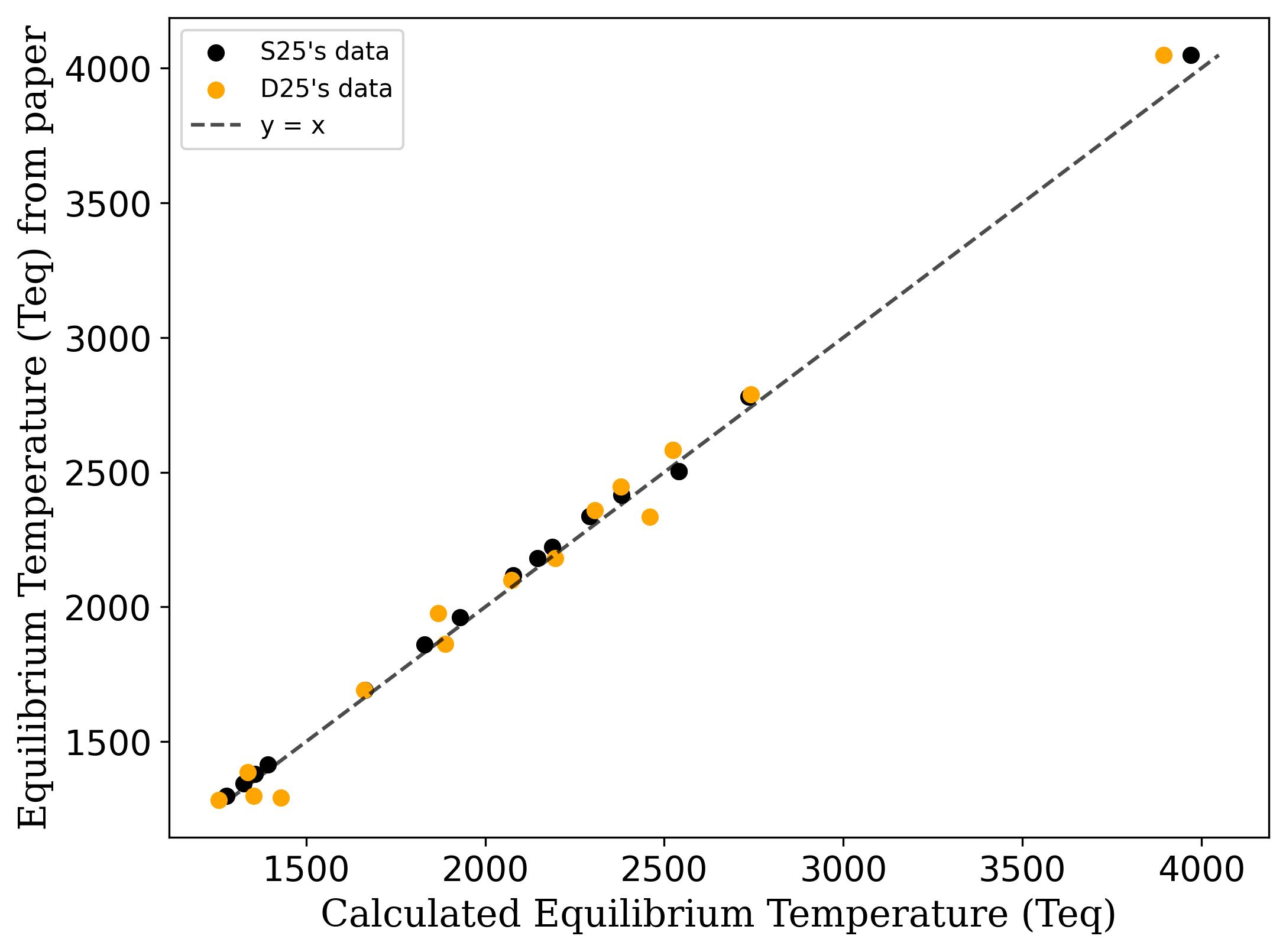}
\hspace{-1in}
\caption{A comparison is made between the equilibrium temperature calculated, using the equation \ref{eq:Teq1} for both S25 and D25, and the equilibrium temperature given in the papers. In orange the data are from D25 and in black they are from S25.}
\label{fig:Temperature}
\end{figure}

As shown in the previous section, for the phase curve parameters, the trends are consistent even if we can see some discrepancies between the studies for individual planets. 
It is interesting to discuss in more detail the results coming from the comparison of the system parameters (figure \ref{fig:comparison_parameters2}). One way to make this comparison is to look at the equilibrium temperature of the planet ($T_{eq}$) from S25 and D25. Using the data from S25 and D25, we can estimate the equilibrium temperature data and see if they are consistent with the equation \ref{eq:Teq1} (figure~\ref{fig:Temperature}) taking $A_{B} = 0$ for S25 and D25.

By recalculating the temperatures using the equation, we can see that they all follow the same trend, with D25 and S25 values close to the unity line meaning that for the equilibrium temperature the data are consistent. Since the equilibrium temperature is not constrained by the Spitzer phase curve, but rather by the choice of stellar and orbital parameters, this demonstrates that the selection of system parameter is not the source of discrepancy. 

So, to further investigate the likely origin of the differences in the results, we now turn to the phase curve parameters, beginning with a discussion of the secondary eclipse depth comparison, where some inconsistencies exceed 2 $\sigma$. As we have seen, two planets have high consistency scores. To compare the secondary eclipse depth we had to transform the S25 representation into the D25 representation using the equation \ref{eq:Fp_Fs}. For that we had to use the value of $\frac{F_{p}}{F_{\star}}$ that can be found in \href{https://excalibur.ipac.caltech.edu/pages/search}{EXCALIBUR pipeline}. 

In this case, a small difference in one parameter leads to a larger difference in the final result, which may explain the 3$\sigma$ difference between D25 and S25 for these two planets.

Another parameter that seems particularly sensitive to variations is offset for which the points are much more scattered than for the other parameters. This parameter corresponds to difference between the substellar point and the actual peak of the brightness curve. As we assume the planets are tidally-locked, this difference can be explained by the presence of a hot spot shifted to a preferred side of the planet due to heat advection, clouds or inhomogeneous composition on the dayside. D25 also experiment with different treatment for detrending detector systematics and found that phase curve offset and amplitude are dependent on the selection of the choice of detrending. As such, this parameter is extremely sensitive to the presence of any small residual systematic errors and can easily vary since phase curve variation sometimes have similar timescale on the variation caused by detector systematics.

Finally, we now turn to consider the results of the period and the semi-major axis. As we saw previously, for all planets outside of WASP-33 b the data have consistency scores below 2$\sigma$.

\begin{figure}[ht]
\includegraphics[scale=0.6]{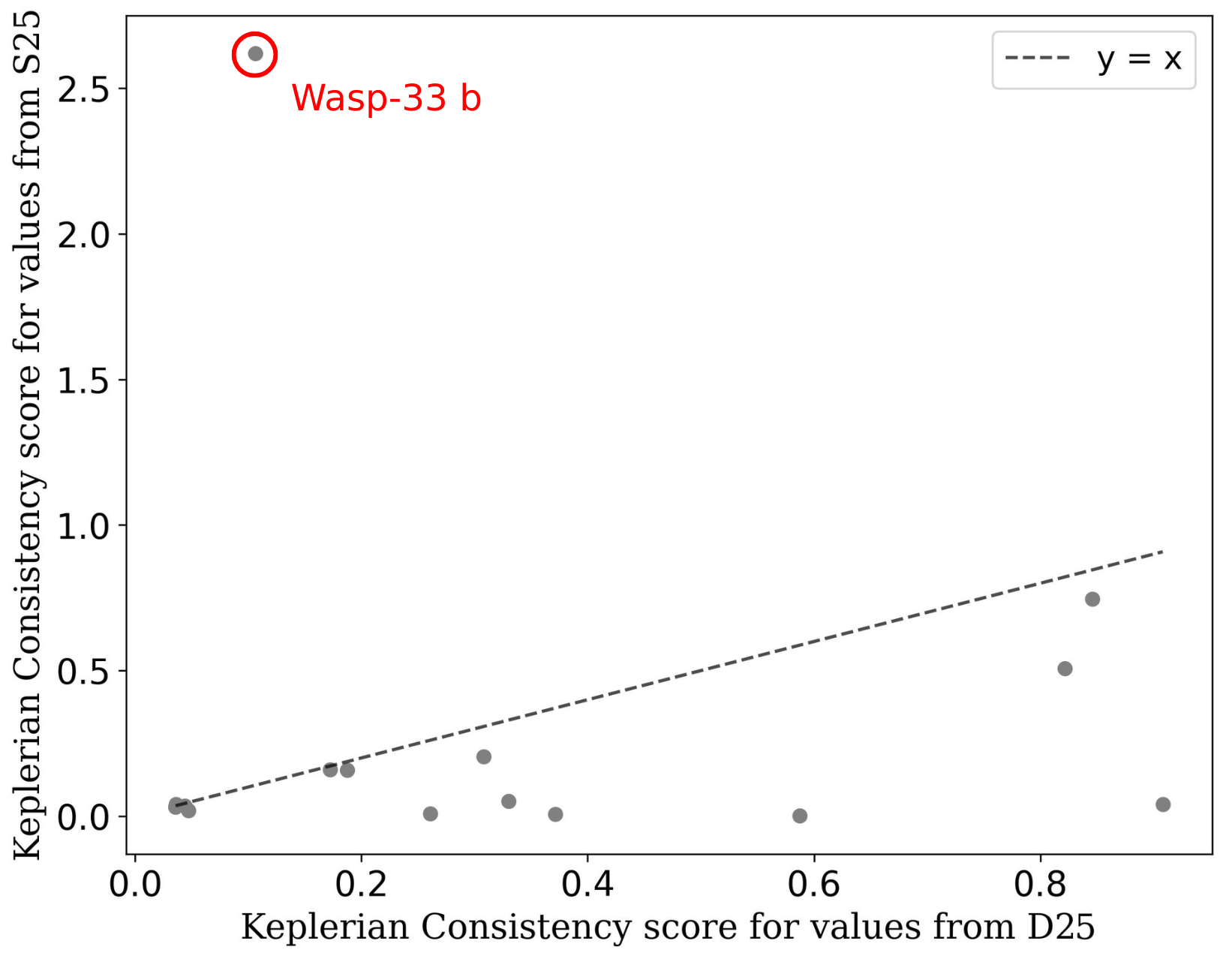}
\hspace{-1in}
\caption{Keplerian Consistency Score (equation \ref{eq:KCS}) between the data from S25 and D25 with all the planets. We have overlaid unity as a dashed line.}
\label{fig:keplerian}
\end{figure}

To see if the data are consistent with the Kepler's law we can now look at the Keplerian Consistency Score. Looking at this score allows us to understand if the inconsistencies are due to system parameter or not. As explained previously the fact that D25 uses an algorithm that selects system parameters based on minimizing the uncertainty and the fact that S25 uses an algorithm which prioritizes identifying a set of self-consistent system parameters could result in some inconsistencies in the system parameters. To see if it's the case or not, we look at the Keplerian Consistency Score, figure~\ref{fig:keplerian}.

Looking at Figure~\ref{fig:keplerian}, we can see, in the data from S25, that there is an outlier point, with a sigma difference of 2.55 (the planet WASP-33 b), which is a significant difference.
The absence of a trend, in this case, suggest that the two studies produce results that are generally consistent with Kepler's laws, with no method yielding better results than the other. However, we only observe the absence of a trend if we remove the point corresponding to WASP-33 b. Since we cannot simply ignore a data point, a more in-depth study of this point is needed to understand why its value is so much higher compared to the others.

\begin{figure}[ht]
\includegraphics[scale=0.17]{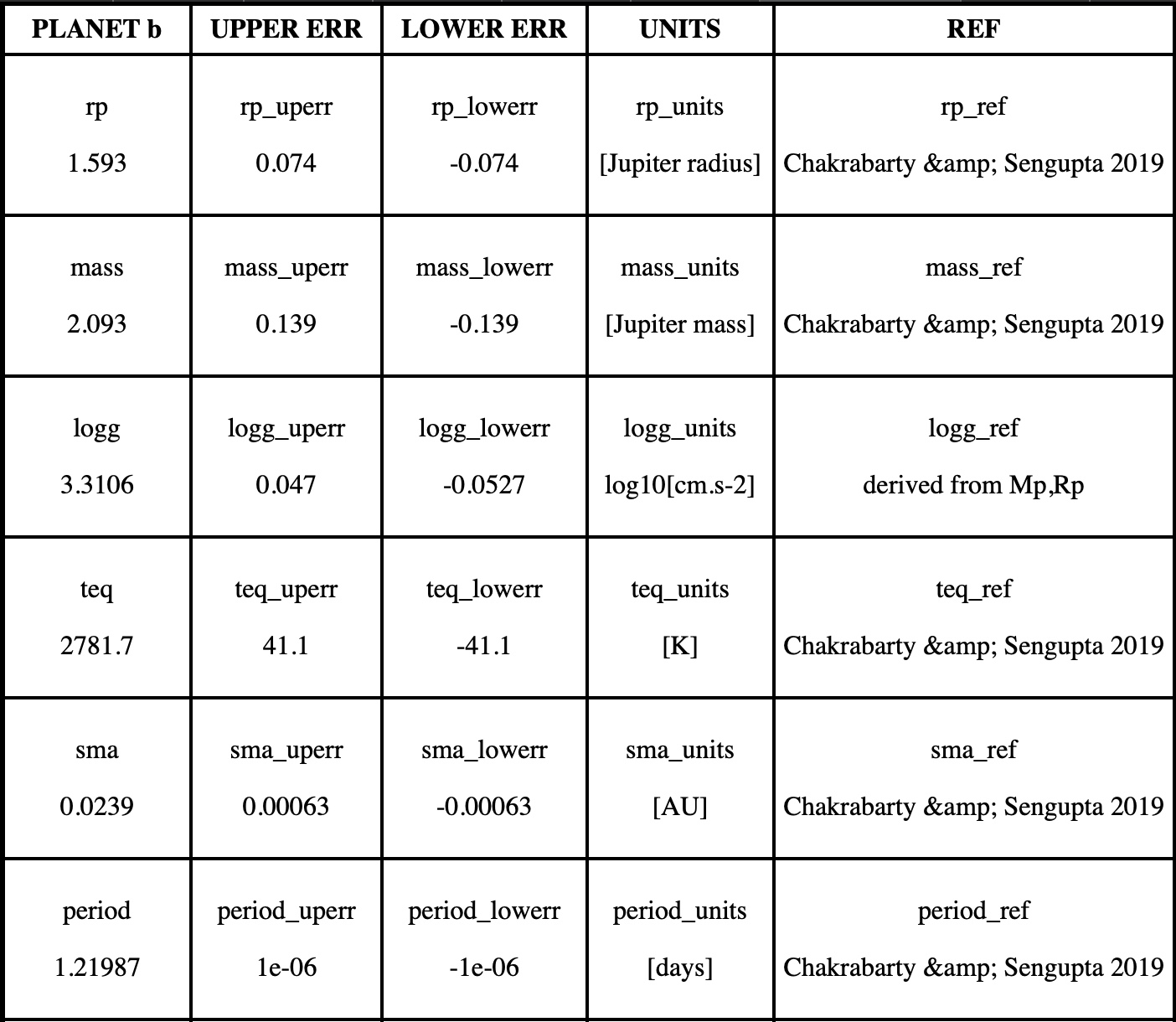}
\hspace{-1in}
\caption{WASP-33 b parameters from  \href{https://excalibur.ipac.caltech.edu/app/db/item?path=698.WASP-33.system.finalize.parameters}{EXCALIBUR pipeline} for the Run ID 698.}
\label{fig:WASP-33b}
\end{figure}

WASP-33 b is a planet that was studied by \citep{Cameron_2010,Johnson_2015,stassun2017,Chakrabarty_2019} over the past decade. To try to understand why the value is so much higher, we can look at the semi-major axis, a parameter that is directly related to the period. D25 used a semi-major axis of 0.0256 AU, 
according to \cite{Cameron_2010}, which is higher than the value used by S25 (0.0239 AU, defined according to \cite{Chakrabarty_2019}). As we saw in figure~\ref{fig:comparison_parameters2}, this difference in the value of semi-major axis for S25 compared with D25 led to a difference of more than 2.5$\sigma$. Knowing this, we need to understand why S25 used this value and so we can look in the EXCALIBUR catalog (figure~\ref{fig:WASP-33b}) to see if it's an algorithm error or something else.

We can see that all the system parameters are self-consistent, as they all come from the same publication \citep{Chakrabarty_2019}. However, if we look at \citep{Chakrabarty_2019} we can see that the physical parameters (such as planet radius, scale parameter and orbital separation) are "\textit{obtained and deduced from their differential transit photometry followed by preprocessing with the wavelet denoising technique and modeling}". In this case the system became physically inconsistent when compared to the Keplerian solution, which explains the discrepancy observed in the semi-major axis.

This highlights that, even if all parameters originate from a single reference, it is prudent to verify the physical consistency of the resulting set.

\begin{figure}[ht]
\centering
\includegraphics[scale=0.58]{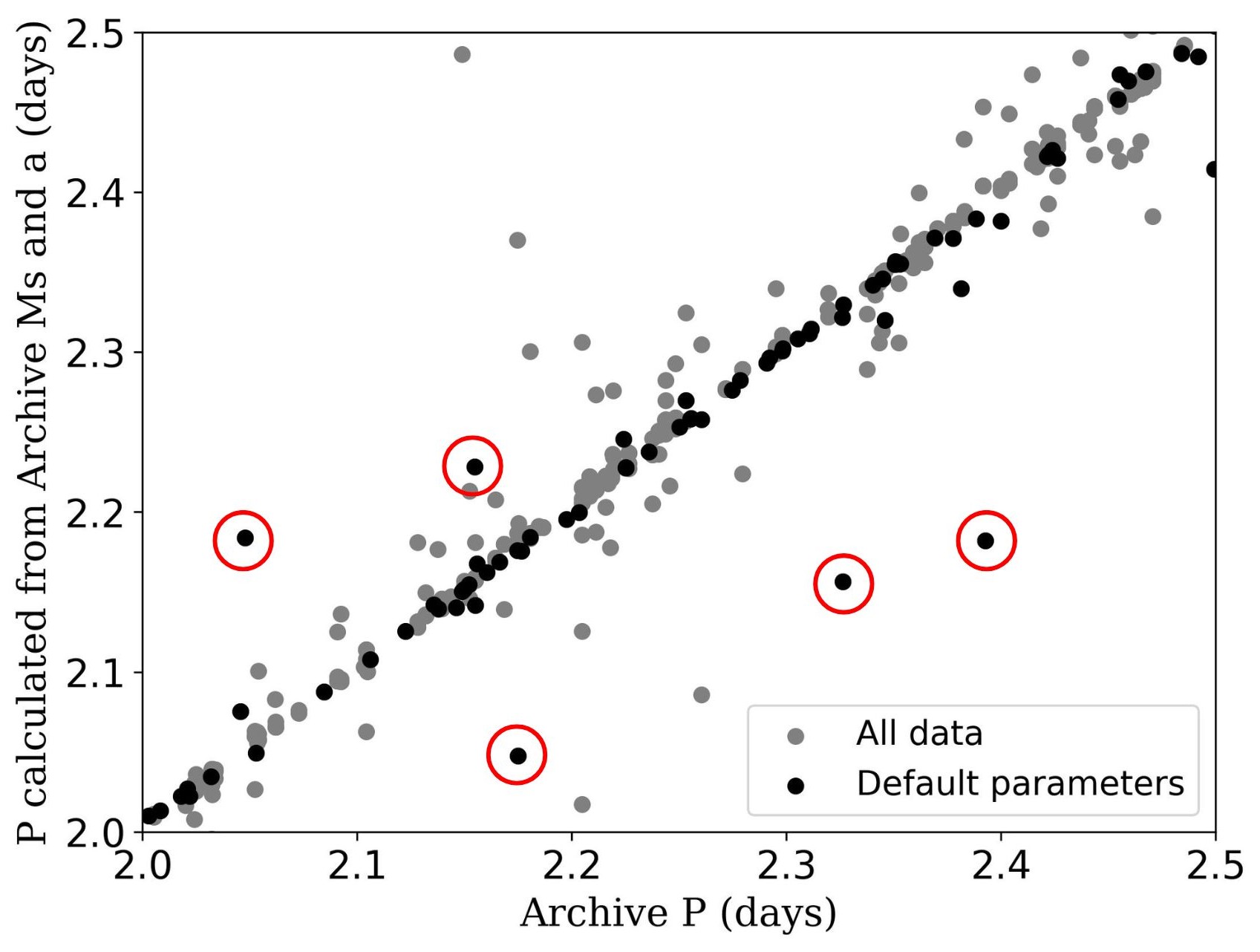}
\hspace{0.5in}
\caption{Comparison between the period computed from the equation \ref{eq:P_Kepler} and the period given in the Exoplanet Archive (for all data in grey and the default parameters in black).
The five outliers highlighted with red circles are discussed in detail below.} 
\label{fig:comparison_parametersEA}
\end{figure}

The fact that there is a possible inconsistency using the same paper leads us to wonder whether there are also possible inconsistencies with \href{https://exoplanetarchive.ipac.caltech.edu/cgi-bin/TblView/nph-tblView?app=ExoTbls&config=PS}{Exoplanet Archive} (NEA). For that we can compute the period using the parameters and plot the results as a function of all the data from NEA and the default parameter from NEA (figure~\ref{fig:comparison_parametersEA}).

We can see that for the NEA ``default parameters", even if most of the points tend to align along the unity slope, for a period between 2 and 2.5 days, we have 5 outliers which are : TOI-6130, TOI-3714, EPIC 201757695, K2-216 and TOI-2266. These outliers shows that NEA parameters are not always consistent with Kepler's law.

The NEA maintains as complete a set of planet parameters as published in the refereed literature.  As a result, there are currently 38,952 planet solutions in the Archive for 6,022 exoplanets in 4,495 planetary systems (obtained on 08 October 2025) - which corresponds to approximately 6 solutions per planet and 9 solutions per system.  The ``default parameters'' for a given planet are just one published solution for that system, but may not represent the ``best'' solution for a given planet. The choice of the ``default parameters'' by the NEA is described on the \href{https://exoplanetarchive.ipac.caltech.edu/docs/faq.html}{FAQ page}. For the 5 outliers identified above, only TOI-3714 b and K2-216b have more than one publications ingested into the Archive. The other three planets only have one refereed publication associated with the planetary parameters. Here we discuss these five specific systems to try to understand why the parameters may not be fully consistent with Kepler's Laws and try to give some possible areas of exploration that might help alleviate the discrepancies.

{\it TOI-6130}: The NEA only has one reference for this planet: the original discovery paper \cite{Ehrhardt2024}.  In that paper, the authors discuss that the activity indicators for TOI-6130 predict a rotation period that is 2$\times$ slower than what is inferred from the combination of $vsini$ and $R_\star$. As a result, the age of the star may be 10Gyr rather than 1Gyr old - as a result, the radius and mass of the star, as reported in the paper and the NEA may be incorrect, and, thus, the resulting planetary mass in the paper may be in error. 

{\it TOI-3714}: The NEA has two papers for TOI-3714; the first paper is the discovery paper \cite{Canas2022} which is also the NEA default parameter set. TOI-3714 appears in a second paper that is primarily a discovery paper for two other TOIs but, in which, the authors present newly derived parameters for TOI-3714 \cite{Hartman2023}. The stellar and planetary parameters between the two data sets are in reasonable agreement with each other except for the stellar metallicity $[Fe/H]$. In the discovery paper, the metallicity is determined to be effectively solar ($[Fe/H]=0.1\pm0.1)$, but in the second paper, the metallicity is significantly not solar ($[Fe/H]=0.39\pm0.09$); additionally, the two papers derive very different ages: 0.7-5 Gyr vs. $12.5\pm6$ Gyr. As a result, the resulting stellar and planetary masses may not be what we think. 

{\it EPIC 201757695}: The NEA only has one paper for this paper planet \cite{delSer2020} which is the original discovery paper. This paper was primarily regarding the statistical validation of transit-like events in the early K2 data using Everest and VESPA. The stellar parameters in the discovery paper were compiled the K2 Ecliptic Plane Input Catalog \citep[EPIC;][]{Huber2016} which derived stellar parameters from photometric surveys. The discovery paper included no new spectroscopic observations of the host star, nor is there any other additional follow-up (e.g., high resolution imaging to identify close-in stellar companions) to characterize the system. As a result, the stellar and planetary parameters for this system may not be as accurate or precise as desired. 

{\it K2-216}: Unlike the other outlier systems, K2-216 has 5 individual papers in the NEA; 2 papers as a candidate planet and 3 papers as a confirmed planet. Two of the confirmed planet papers are general K2 survey papers \citep{Livingston2018, Mayo2018} while the third is a paper dedicated to K2-216b specifically \citep{Persson2018}, which is the ``default'' paper in the NEA. Overall, the three papers agree reasonably well with each other regarding the stellar and planetary parameters. However, the stellar parameters for K2-216 are derived from a variety of sources. The stellar mass is obtained from spectral fitting using SpecMatch-Emp/Torres and BASTA; the stellar radius is obtained from the Stefan-Boltzmann and the derived effective temperature, Gaia distance, and apparent visual magnitude; the stellar density is derived from the transit fitting model. As a result, the stellar parameters themselves may not be fully consistent with each other. For example, the stellar mass of $0.7 M_\odot$ and the stellar radius of $0.72 R_\odot$ imply a surface gravity of $\rho = 2.7\ g/cm^3$ but the paper reports a density of $\rho = 2.3\ g/cm^3$. While the uncertainties listed in the Table 5 in the paper may be sufficient to yield an overall assessment of the star and the planetary system, the disagreement between with derived Kepler-based period may indicate that the stellar parameters and, hence, the planet parameters are biased.

{\it TOI-2266}: The NEA archive only has the discovery paper for TOI-2266 b \citep{Parviainen2024}. In this paper, the authors only obtain radius for the planet based upon the transit depth -- no radial velocity detections of the planet or its orbit were made.  Thus, the planet is only statistically validated and the mass and orbit parameters (e.g., eccentricity) of the planet are unknown. The planet parameters are derived from the multi-colour transit fit analysis. The disagreement with the derived Kepler-based period may indicate that the system parameters may be biased. 

One needs to be careful to assign a source to observed discrepancies. Each of these systems would require more inquiry and perhaps additional observations to understand in detail from where the discrepancies may lie. These results simply serve as a reminder that the self-consistency amongst the stellar, planet, and orbital parameters is key to understanding the systems in detail and taken as a group.

\section{Conclusions}\label{sec:Conclusion}

In this study, we conducted an initial comparison of the results from catalogues of the two most recent phase curve studies. As we have seen, these two studies have four significant differences: 1) they use different methods to model instrument systematics, 2) they offer different interpretations of how uniform analysis is defined, 3) they use different approaches to select the optimal parameters of the exoplanetary system from among numerous measurements, and finally, 4) they focus on different target samples. D25 focuses on phase curve trends at 4.5 $\mu$m, while S25 focuses on phase curve trends at 3.6 $\mu$m and 4.5 $\mu$m. We therefore compared the D25 and S25 studies in this study for 4.5 $\mu$m only. To do this, we looked at the consistency scores between the different parameters characterizing the phase curves to see if there were any significant differences.

This study shows that the results are generally consistent despite the presence of a few differences, mainly in the phase curve offset. For population level types of analysis, we find the D25 and S25 methods give consistent results. In order to explain the origin of the differences between the D25 and S25 methods at the individual planet level, we compared the system parameters between the two studies (the semi-major axis, the period and the equilibrium temperature) and found cases where the system parameters do not satisfy Kepler's laws. We investigated the system parameter in the NASA Exoplanet Archive and found that even when the planet system parameters are taken from the same publication, they do not necessarily comply with Kepler's laws (as we saw with WASP-33 b and the five outliers). We conclude it is therefore prudent to check for Keplerian consistency when selecting exoplanet system parameters.

\begin{acknowledgements}
This research has made use of the NASA Exoplanet Archive, which is operated by the California Institute of Technology, under contract with the National Aeronautics and Space Administration under the Exoplanet Exploration Program.
The research was carried out at the Jet Propulsion Laboratory, California Institute of Technology, under a contract with the National Aeronautics and Space Administration (80NM0018D0004)
\end{acknowledgements}

\section*{Declaration}

The LLM "DeepL Translator" was used for spell checking and grammar correction of certain sentences in this work. All ideas, methods, and results are solely the work of the authors.

\clearpage
\appendix

In this appendix, we present several tables the first one gathers the various constants that have been used in this study, and the second one lists the values of the four key parameters; the planet-to-star radius ratio $\frac{R_{P}}{R_{\star}}$, the eclipse depth $\frac{F_{P}}{F_{\star}}$, the phase curve amplitude, and the phase curve offset, which are reported in S25 and D25.
\vspace{1.5em} 
\section{Constants}

\begin{table}[H]
\caption{Descriptive version of the constants used in the paper.}

\label{tab:constants}
\begin{center}
\begin{tabular}{lll} \hline \hline
 Parameter & Value & Units \\ \hline
 1 AU & $1.496 \times 10^{11}$ & $\rm m$ \\ 
 \text{$R_{J}$} & $6.9911 \times 10^{7}$  & $\rm m $ \\
  \text{$R_{\odot}$} & $6.957 \times 10^{8}$  & $\rm m $  \\
  \text{$M_{\odot}$} &  $1.989 \times 10^{30}$  & $\rm kg $ \\
  \text{$L_{\odot}$} &  $3.828 \times 10^{26}$  & $\rm W $ \\
  \text{Gravitational constant (G)} & $6.67 \times 10 ^{-11}$ & $\rm m^{3}~kg^{-1}~s^{-2}$ \\
  \text{Stefan-Boltzmann constant} ($\sigma$) & $5.67 \times 10^{-8}$ & $\rm W~m^{-2}~K^{-4}$ \\
\end{tabular}
\\
\end{center}
\end{table}

\section{Phase curve Parameters Data for Both Studies}

\begin{table}[H]
\caption{Descriptive version table of the phase curve parameters.}
\label{tab:description}
\begin{center}
\begin{tabular}{rllllllll} \hline \hline
Planet & $\frac{R_{p}}{R_{\star }}_{D25}$ &$\frac{R_{p}}{R_{\star }}_{S25}$ & $\frac{F_{p}}{F_{\star }}_{D25}$ & $\frac{F_{p}}{F_{\star }}_{S25}$ & $A_{D25}$ & $A_{S25}$ & $\phi_{D25}$ & $\phi_{S25}$ \\ \hline

Qatar-2 b 
& $0.165 \substack{+0.0002 \\ -0.0002}$\ 
& $0.160 \substack{+0.0024 \\ -0.0024}$\ 
& $3100 \substack{+300 \\ -300}$\ 
& $2907.25 \substack{+371.37 \\ -371.37}$\ 
& $1000 \substack{+403.11 \\ -403.11}$\ 
& $ 1205 \substack{+53.1 \\ -53.1}$\ 
& $ 30 \substack{+20 \\ -20}$\ 
& $-18 \substack{+0.8 \\ -0.8}$\  \\

WASP-43 b 
& $0.160 \substack{+0.001 \\ -0.001}$\ 
& $0.16 \substack{+0.014 \\ -0.018}$\ 
& $3800 \substack{+100 \\ -100}$\ 
& $3594.53 \substack{+190.66 \\ -190.66}$\  
& $1850 \substack{+70.71 \\ -70.71}$\ 
& $ 1638 \substack{+42.4 \\ -42.4}$\ 
& $  9 \substack{+3 \\ -3}$\ 
& $-1 \substack{+0.1 \\ -0.1}$\  \\

WASP-52 b 
& $0.16 \substack{+0.002 \\ -0.002}$\ 
&  $0.16 \substack{+0.0056 \\ -0.00556}$\ 
& $3500 \substack{+300 \\ -300}$\ 
& $2873.76 \substack{+161.85 \\ -161.85}$\  
& $1750 \substack{+430.12 \\ -430.12}$\ 
& $ 479 \substack{+22.3 \\ -22.3}$\ 
& $ 60 \substack{+10 \\ -20}$\ 
& $51 \substack{+3.3 \\ -3.3}$\ \\

Qatar-1 b 
&  $0.15 \substack{+0.002 \\ -0.002}$\ 
& $0.14 \substack{+0.0052 \\ -0.0052}$\ 
& $3100 \substack{+300 \\ -300}$\ 
& $2602.09 \substack{+169.93 \\ -169.93}$\ 
& $1350 \substack{+291.54 \\ -291.54}$\ 
& $ 1132 \substack{+84.3 \\ -84.3}$\ 
& $-30 \substack{+20 \\ -20}$\ 
& $-9 \substack{+0.7 \\ -0.7}$\ \\

WASP-77 AB 
&  $0.11 \substack{+0.0007 \\ -0.0008}$\
& $0.13 \substack{+0.0055 \\ -0.0044}$\  
& $2180 \substack{+90 \\ -90}$\ 
& $2616.59 \substack{+224.57 \\ -224.57 }$\ 
& $745 \substack{+109.66 \\ -109.66}$\ 
& $ 860 \substack{+24.8 \\ -24.8 }$\ 
& $14 \substack{+8 \\ -8}$\ 
& $18 \substack{+0.5 \\ -0.5}$\ \\

WASP-14 b 
& $0.095 \substack{+0.0006 \\ -0.0006}$\ 
& $0.098 \substack{+0.0080 \\ -0.0080}$\ 
& $2540 \substack{+80 \\ -80}$\ 
& $2275.55 \substack{+136.06 \\ -136.06}$\ 
& $685 \substack{+56.57 \\ -56.57}$\ 
& $ 820 \substack{+22.1 \\ -22.1}$\
& $2 \substack{+4 \\ -4}$\ 
& $11 \substack{+0.3 \\ -0.3}$\ \\

KELT-14 b
& $0.112 \substack{+0.002 \\ -0.002}$\ 
&  $0.114 \substack{+0.0040 \\ -0.0040}$\  
& $2900 \substack{+200 \\ -100}$\
& $2866.61 \substack{+215.08\\ -215.08}$\ 
& $500 \substack{+180.28\\ -180.28}$\ 
& $ 869 \substack{+78.5 \\ -78.5}$\
& $60 \substack{+10 \\ -20}$\
& $12 \substack{+1.1 \\ -1.1}$\  \\

WASP-19 b 
& $0.139 \substack{+0.002 \\ -0.002}$\
& $0.14 \substack{+0.0057 \\ -0.0065}$\ 
& $5500 \substack{+300 \\ -300}$\ 
& $5184.89 \substack{+250.19 \\ -250.19}$\ 
& $2250 \substack{+250\\ -250}$\ 
& $ 1608 \substack{+89.1 \\ -89.1}$\
& $-17 \substack{+4 \\ -4}$\
& $7 \substack{+0.4 \\ -0.4}$\ \\

WASP-76 b
& $0.105 \substack{+0.0008 \\ -0.0009}$\
& $0.106 \substack{+0.0042 \\ -0.0034}$\ 
& $3400 \substack{+100 \\ -100}$\ 
& $3286.08 \substack{+166.17 \\ -166.17}$\  
& $1250 \substack{+158.11\\ -158.11}$\ 
& $ 1309 \substack{+59.9 \\ -59.9}$\ 
& $-3 \substack{+7 \\ -9}$\ 
& $11 \substack{+0.5 \\ -0.5}$\ \\

WASP-121 b
& $0.125 \substack{+0.0009 \\ -0.0009}$\ 
& $0.12 \substack{+0.0037 \\ -0.0037}$\  
&$4800 \substack{+100 \\ -100}$\ 
& $4586.79 \substack{+807.77 \\ -807.77}$\ 
& $2150 \substack{+206.16\\ -206.16}$\
& $ 2037 \substack{+78.5 \\ -78.5}$\ 
& $-5 \substack{+4 \\ -4}$\ 
& $-1 \substack{+0 \\ -0}$\  \\

HAT-P-7 b 
& $0.077 \substack{+0.001 \\ -0.001}$\
& $0.075 \substack{+0.015 \\ -0.015}$\  
& $2200 \substack{+100 \\ -100}$\ 
& $1969.52 \substack{+116.75 \\ -116.75}$\ 
& $450 \substack{+180.28\\ -180.28}$\ 
& $ 878 \substack{+53.3 \\ -53.3}$\ 
& $-60 \substack{+20 \\ -20}$\ 
& $-17 \substack{+1.1 \\ -1.1}$\ \\

KELT-1 b 
& $0.076 \substack{+0.001 \\ -0.001}$\ 
& $0.076 \substack{+0.0029 \\ -0.027}$\ 
& $2300 \substack{+100 \\ -100}$\ 
& $2159.99 \substack{+578.56 \\ -578.56}$\ 
& $900 \substack{+111.80\\ -111.80}$\ 
& $ 717 \substack{+51.2 \\ -51.2}$\ 
& $5 \substack{+5 \\ -5}$\ 
& $2 \substack{+0.1 \\ -0.1}$\  \\

WASP-12 b (v1) 
& $0.108 \substack{+0.002 \\ -0.001}$\ 
& $0.12 \substack{+0.014 \\ -0.014}$\ 
& $4100 \substack{+200 \\ -200}$\ 
& $205.68 \substack{+101.30 \\ -101.30}$\ 
& $2150 \substack{+250\\ -250}$\ 
& $ 1474 \substack{+73 \\ -73}$\ 
& $-8 \substack{+5 \\ -4}$\ 
& $2 \substack{+0.1 \\ -0.1}$\  \\

WASP-12 b (v2) 
& $0.105 \substack{+0.001 \\ -0.001}$\  
& $0.12 \substack{+0.014 \\ -0.014}$\ 
& $4200 \substack{+200 \\ -200}$\ 
& $205.68 \substack{+101.30 \\ -101.30}$\ 
& $2150 \substack{+250\\ -250}$\ 
& $ 1474 \substack{+73 \\ -73}$\ 
& $30 \substack{+8 \\ -8}$\ 
& $2 \substack{+0.1 \\ -0.1}$\  \\

WASP-33 b 
& $0.11 \substack{+0.0005 \\ -0.0005}$\ 
&  $0.11 \substack{+0.005 \\ -0.005}$\ 
& $4430 \substack{+60 \\ -60}$\ 
& $4254.58 \substack{+190.95 \\ -190.95}$\ 
& $675 \substack{+51.48\\ -51.48}$\ 
& $ 1810 \substack{+10.5 \\ -10.5}$\ 
& $12 \substack{+1 \\ -1}$\ 
& $17 \substack{+0.1 \\ -0.1}$\  \\

KELT-9 b 
& $0.081 \substack{+0.0006 \\ -0.0005}$\ 
& $0.080 \substack{+0.004 \\ -0.003}$\ 
& $2890 \substack{+40 \\ -40}$\ 
& $2730.23 \substack{+20.51 \\ -20.51}$\  
& $1870 \substack{+35.36\\ -35.36}$\ 
& $ 838 \substack{+18.4 \\ -18.4}$\ 
& $49 \substack{+4 \\ -4}$\ 
& $27 \substack{+0.6 \\ -0.6}$\  \\
\end{tabular}
\\ 
\end{center}
\end{table}

\vspace{1.5em} 

\section{Mathematical Results Used to Generate Figures in the Results Section.}

We now present three tables showing the mathematical results used to generate Figures \ref{fig:phys_parms}, \ref{fig:comparison_parameters}, and \ref{fig:comparison_parameters2} in the Results section.

\setlength{\textfloatsep}{2pt}
\setlength{\floatsep}{2pt}  
\setlength{\intextsep}{2pt}

\begin{table}[H]
\caption{Table listing slopes and y-intercepts with associated uncertainties obtained from Orthogonal Distance Regression and Ordinary Least Squares fits, showed in figure \ref{fig:phys_parms}, comparing the results from the two studies.} \label{tab:slope}
\begin{center}
\begin{tabular}{l|ll|ll} \hline \hline
  & \multicolumn{2}{c}{ODR} &  \multicolumn{2}{c}{LDS} \\
 Parameter & slope & intercept & slope & intercept  \\
 \hline
 $\frac{R_{p}}{R_{\star }}$ & $1.055 \substack{+0.038 \\ -0.038}$ & $-0.006 \substack{+0.005 \\ -0.005}$ & 1.006 & $-$0.003\\ 
 Eclipse Depth & $0.878 \substack{+0.099 \\ -0.099}$ & $479.729 \substack{+283.03 \\ -283.03}$ & 0.730 & 934.067 \\  
  Amplitude & $1.228 \substack{+0.102 \\ -0.102}$ & $-296.537 \substack{+145.204 \\ -145.204}$ & 1.102 & 51.03\\ 
  Offset & $1.019 \substack{+0.329 \\ -0.329}$ & $-3.132 \substack{+5.175 \\ -5.175}$ & 1.211 & 0.623 \\  
\end{tabular}
\\
\end{center}
\end{table}


\begin{table}[H]
\caption{Table listing consistency scores, showed in figure \ref{fig:comparison_parameters}, for each of the four parameters ($\frac{R_{p}}{R_{\star }}$, $Depth_{Eclipse}$, A and $\phi$).} \label{tab:comp4}
\begin{center}
\begin{tabular}{l|llll} \hline \hline
 
Planet & $\frac{R_{p}}{R_{\star }}$ & $Depth_{Eclipse}$ & $A$ & $\phi$ \\ \hline

Qatar-2 b & 1.89 & 0.40 & 0.50 & 2.40  \\

WASP-43 b 
& 0.23 & 0.95 & 2.57 & 3.33 \\

WASP-52 b 
& 0.77 & 1.84 & 2.95 & 0.85 \\

Qatar-1 b 
&  0.66 & 1.44 & 0.72 & 1.05 \\

WASP-77 AB 
& 1.33 & 1.80 & 1.02 & 0.50 \\

WASP-14 b 
& 0.48 & 1.68 & 2.22 & 2.24 \\

KELT-14 b
& 0.52 & 0.11 & 1.88 & 4.77 \\

WASP-19 b & 0.21 & 0.81 & 2.42 & 5.97  \\

WASP-76 b & 0.22 & 0.59 & 0.35 & 1.99 \\

WASP-121 b & 1.37 & 0.26 & 0.51 & 1.00 \\

HAT-P-7 b & 0.11 & 1.50 & 2.28 & 2.15 \\

KELT-1 b & 0.003 & 0.24 & 1.49 & 0.75 \\

WASP-12 b (v1) & 0.96 & 3.06 & 2.60 & 2.00 \\

WASP-12 b (v2) & 1.18 & 3.30 & 4.50 & 3.50  \\

WASP-33 b & 0.03 & 0.58 & 1.63 & 4.98 \\

KELT-9 b & 0.21 & 3.55 & 2.98 & 5.44 \\
\end{tabular}
\\
\end{center}
\end{table}

\begin{table}[H]
\caption{Table listing consistency scores, showed in figure \ref{fig:comparison_parameters2}, for each of the four parameters, Period$_{S25}$, Period$_{D25}$, Semi-major axis and $T_{eq}$.}
\label{tab:comp3}
\begin{center}
\begin{tabular}{l|llll} \hline \hline

Planet & Period$_{S25}$ & Period$_{D25}$ & Semi-major axis & $T_{eq}$ \\ \hline

Qatar-2 b & 0.16 & 0.19 & 0.11 & 2  \\

WASP-43 b 
& 0.04 & 0.04 & 0.0 & 0.82 \\

WASP-52 b 
& 0.16 & 0.17 & 0.0 & 0.0 \\

Qatar-1 b 
&  0.02 & 0.05 & 0.04 & 0.52 \\

WASP-77 AB 
& 0.20 & 0.31 & 0.08 & 0.0 \\

WASP-14 b 
& 0.001 & 0.59 & 0.72 & 0.003 \\

KELT-14 b
& 0.04 & 0.91 & 0.89 & 0.24 \\

WASP-19 b 
& 0.05 & 0.33 & 0.41 & 0.12  \\

WASP-76 b 
& 0.03 & 0.04 & 0.0 & 0.0 \\

WASP-121 b 
& 0.75 & 0.85 & 0.07 & 0.009 \\

HAT-P-7 b 
& 0.006 & 0.37 & 0.35 & 0.20 \\

KELT-1 b 
& 0.51 & 0.82 & 0.0 & 0.124 \\

WASP-12 b (v1) 
& 0.03 & 0.0 & 0.0 & 0.41 \\

WASP-12 b (v2) 
& 0.03 & 0.0 & 0.0 & 0.41  \\

WASP-33 b 
& 2.62 & 0.11 & 2.57 & 0.05 \\

KELT-9 b 
& 0.008 & 0.26 & 0.26 & 0.003 \\ 
\end{tabular}
\\
\end{center}
\end{table}

\clearpage

\bibliography{ref}{}
\bibliographystyle{aasjournalv7}

\end{document}